\documentclass{jfm}
\usepackage{graphicx}
\usepackage{epstopdf, epsfig}
\usepackage{dcolumn}
\usepackage{bm}
\usepackage{amsfonts}
\usepackage{amssymb}
\usepackage{amsmath}    
\usepackage{graphicx}   
\usepackage{verbatim}   
\usepackage{color}      
\usepackage[colorlinks=true,linkcolor=blue,citecolor=blue,allcolors=blue]{hyperref}

\shorttitle{On the stability of inhomogeneous fluids under acoustic fields}
\shortauthor{V.K.Rajendran, S.P.Aravind Ram and K.Subramani}

\title{On the stability of inhomogeneous fluids under acoustic fields}

\author{Varun Kumar Rajendran\aff{1},
  Aravind Ram S P
 \and Karthick Subramani\aff{1}
 \corresp{\email{karthick@iiitdm.ac.in}}}

\affiliation{\aff{1}Department of Mechanical Engineering, Indian Institute of Information Technology, Design and Manufacturing, Kancheepuram, Chennai-600127, India}

\begin{document}

\maketitle

\begin{abstract}
In this work, we present the stability theory for inhomogeneous fluids subjected to standing acoustic fields. Starting from the first principles, the stability criterion is established for two fluids of different acoustic impedance separated by a plane interface. Through stability theory and numerical simulations we show that,
in the presence of interfacial tension, the relocation of high-impedance fluid from anti-node to node occurs when the acoustic force overcomes interfacial tension force, which is in agreement with recent microchannel experiments. Furthermore, we establish an acoustic Bond number that characterizes stable ($Bo_{a}<1$)  and relocation ($Bo_{a}>1$) regimes. Remarkably, it is found that the critical acoustic energy density required for relocation can be significantly reduced by increasing the channel height which could help design acoustofluidic microchannel devices that handle immiscible fluids.
\end{abstract}

\section{Introduction}\label{Sec 1}
When an acoustic field encounters inhomogeneity, it exerts acoustic radiation force on it. Here by inhomogeneity, we mean non-uniform or discontinuous variation of physical properties in a system such as particles/cells suspended in fluid, emulsions, co-flowing streams of miscible or immiscible fluids, and fluid subjected to a temperature gradient. The acoustic forces acting on inhomogeneity are extensively studied in microscale flows, and this field is known as 'microscale acoustofluidics'\citep{Friend2011Jun}. Over the last two decades, acoustofluidics has found a wide range of applications in biological\citep{Ahmed2016Mar,Iranmanesh2015Sep,Collins2015Nov,Christakou2013Apr,Lakshmanan2020Sep}, chemical\citep{Shi2009Dec,Xie2020Dec}, and medical\citep{Li2015Apr,Lu2019Feb,Zhang2020Mar} sciences. 

Recently, the relocation and stabilization of inhomogeneous co-flowing fluid streams in microchannels has gained the attention of the research community which is evident from the following works. Through silicon-glass microchannel experiments, \citet{Deshmukh2014Jul} could relocate high-impedance sodium chloride solution to node (center) and low-impedance water to anti-node (sides). Also, they could stabilize high-impedance fluid at the center (and low-impedance fluid to the sides) against gravity stratification using acoustic forces. In addition to the above experiments on miscible fluids, \citet{Hemachandran2019Apr} demonstrated the relocation of immiscible fluids using acoustic fields by overcoming the interfacial tension forces. Followed by this, \citet{Karlsen2018Jan} showed that acoustic forces acting on stable inhomogeneous fluid configuration could effectively suppress the boundary-driven Rayleigh streaming in the bulk. The theoretical framework and understanding of the above non-linear acoustic forces on inhomogeneous fluids are provided by \citet{Rajendran2022Jun,Karlsen2016Sep,Karlsen2018Jan}. Other notable works on the practical applications of acoustic forces on co-flowing inhomogeneous fluids include iso-acoustic focusing of cells \citep{augustsson2016iso}, acoustic focusing of sub-micron particles \citep{VanAssche2020Feb,Gautam2018Oct}, tweezing and patterning of inhomogeneous fluids in a microchannel \citep{Karlsen2017Mar,Baudoin2020Aug}, rapid mixing of fluids using an alternating multinode method \citep{Pothuri2019Dec}, reversible stream-droplet transition in a microfluidic co-flowing immiscible system \citep{Hemachandran2021Sep}. Despite the above recent advancements and practical importance, the criterion at which the inhomogeneous fluid system becomes unstable or stable under acoustic fields has not been clearly established. This paper aims to establish the stability criterion of inhomogeneous co-flowing fluids subjected to standing acoustic wave fields.  

In this work, using linear stability analysis, we derive the dispersion relation that governs the stability of inhomogeneous fluids (with and without interfacial tension) under acoustic body force. We study the various parameters such as the initial arrangement of fluids, the position of the interface with respect to the node, acoustic energy density, the height of the channel, and surface tension to establish the necessary and sufficient conditions for relocation and stability. For fluids with interfacial tension, a non-dimensional number called acoustic Bond number is obtained theoretically which characterizes stable and unstable (relocation) regime. Also, we deduce a relation between critical acoustic energy density and the height of the channel which paves a way for relocating fluids with higher interfacial tension $(\mathcal{O}(10^1$ to $10^2) mN/m)$ in a microchannel. Furthermore, numerical simulations are carried out using generalized acoustic body force which agrees well with the derived theoretical stability criterion.  

\section{Physics of the problem}\label{Sec 2}
The hydrodynamics of the inhomogeneous fluids involved in this study is governed by the mass-continuity and momentum equations \citep{Landau1987Aug},
\begin{subequations}
\label{eq 1}
    \begin{gather}
    \label{eq 1a}
        \frac{\partial\rho}{\partial t}+\boldsymbol{\nabla\cdot}(\rho \textbf{\emph{V}})=0, \\
    \label{eq 1b}
        \rho \frac{D\textbf{\emph{V}}}{Dt}= -\boldsymbol{\nabla} P +\eta\nabla^2 \textbf{\emph{V}} +\beta\eta\boldsymbol{\nabla}(\boldsymbol{\nabla\cdot}\textbf{\emph{V}}) + \textbf{\emph{f}}_{ac}.
    \end{gather}
\end{subequations}
where $\rho$ represents density, $\textbf{\emph{V}}$ represents the velocity vector field, $P$ represents the pressure field, $\eta$ is the dynamic viscosity of the fluid,  $\beta=(\xi/\eta)+(1/3)$, $\xi$ is the bulk viscosity, and $D/Dt$ denotes the material derivative ($D/Dt=\partial_t+\textbf{V}\boldsymbol{\cdot\nabla}$). Here the body force $\textbf{\emph{f}}_{ac}$ is only due to acoustics. The gravitational body force is neglected since it is dominated by acoustic force in microscale flows. The acoustic body force $\textbf{\emph{f}}_{ac}$ is given as \citep{Rajendran2022Jun}
\begin{align}
\label{eq 2}
     \textbf{\emph{f}}_{ac} = -\boldsymbol{\nabla\cdot}\langle\rho_0\textbf{\emph{v}}_1 \textbf{\emph{v}}_1\rangle = & \left( \frac{1}{2}\boldsymbol{\nabla}\left(\kappa_0\langle|p_1|^2\rangle - \rho_0 \langle|\textbf{\emph{v}}_1|^2\right\rangle)\right) + \biggl[\langle\textbf{\emph{v}}_1\times\boldsymbol{\nabla}\times(\rho_0\textbf{\emph{v}}_1)\rangle\biggr] \nonumber \\ 
      & + \left(-\frac{1}{2}\langle|p_1|^2\rangle\boldsymbol{\nabla}\kappa_0-\frac{1}{2}\langle\textbf{\emph{v}}_1^2\rangle\boldsymbol{\nabla}\rho_0\right)  \\   
      \label{eq 3} 
            =& \ (\textbf{\emph{f}}_{1})+[\textbf{\emph{f}}_{2}]+(\textbf{\emph{f}}_{3}).
\end{align}
where $p_1$ and $\textbf{\emph{v}}_1$ denote the first-order (fast time scale) pressure and velocity fields due to acoustic waves (see Appendix \ref{Sec A1}) and $\langle...\rangle$ is the time average in one oscillation period (the time average of two first-order fields $\langle\textbf{\emph{u}}_1\textbf{\emph{v}}_1\rangle$ is defined as $\frac{1}{2}$Real($\textbf{\emph{u}}_1^{\star}\textbf{\emph{v}}_1$), where $\star$ denotes complex conjugation). The terms $\rho_0$ and $\kappa_0$ denote the zeroth-order (background) density and compressibility of the fluid. 
In (\ref{eq 2}), the first term is a conservative or gradient term that induces pressure and not fluid flow, the second term is only dominant at boundary layers and is responsible for boundary-driven Rayleigh streaming and the third term is responsible for relocation and stabilization of inhomogeneous fluids. Hence, only the relevant third term ($\textbf{\emph{f}}_{3}$ in (\ref{eq 3})) is considered for theoretical analysis. For the standing acoustic wave applied along the X-direction, $\langle|p_1|^2\rangle=p_a^2\sin^2(k_wx)$, $\langle|\textbf{\emph{v}}_1^2|\rangle=p_a^2/(\rho_0^2c_0^2)\cos^2(k_wx)$, $k_w = 2\upi/\lambda_w$ denotes the wavenumber ($\lambda_w$ denotes the wavelength) and $p_a$ denotes the pressure amplitude. Then the relocation force $f_{rl}$ can be well approximated in terms of impedance gradient as \citep{Rajendran2022Jun}
\begin{equation}
\label{eq 4}
    \textbf{\emph{f}}_{3}=\textbf{\emph{f}}_{rl}=-E_{ac}\cos({2k_wx})\boldsymbol{\nabla}\hat Z.
\end{equation}
where $E_{ac}=p_a^2/(4\rho_{avg}c_{avg}^2)=(v_a^2\rho_{avg})/4$ is the acoustic energy density, $ Z = \rho_0 c_0$ denotes impedance, $c_0$ denotes background (zeroth-order) speed of sound in a medium, $\hat Z=Z/z_{avg}$, $\hat c_0 = c_0/ c_{avg}$ and $\hat \rho_0 = \rho_0/ \rho_{avg}$, where the subscript 'avg' denotes the respective average quantities of fluid A and B.
\begin{figure}
    \center
    \includegraphics[width=1\linewidth]{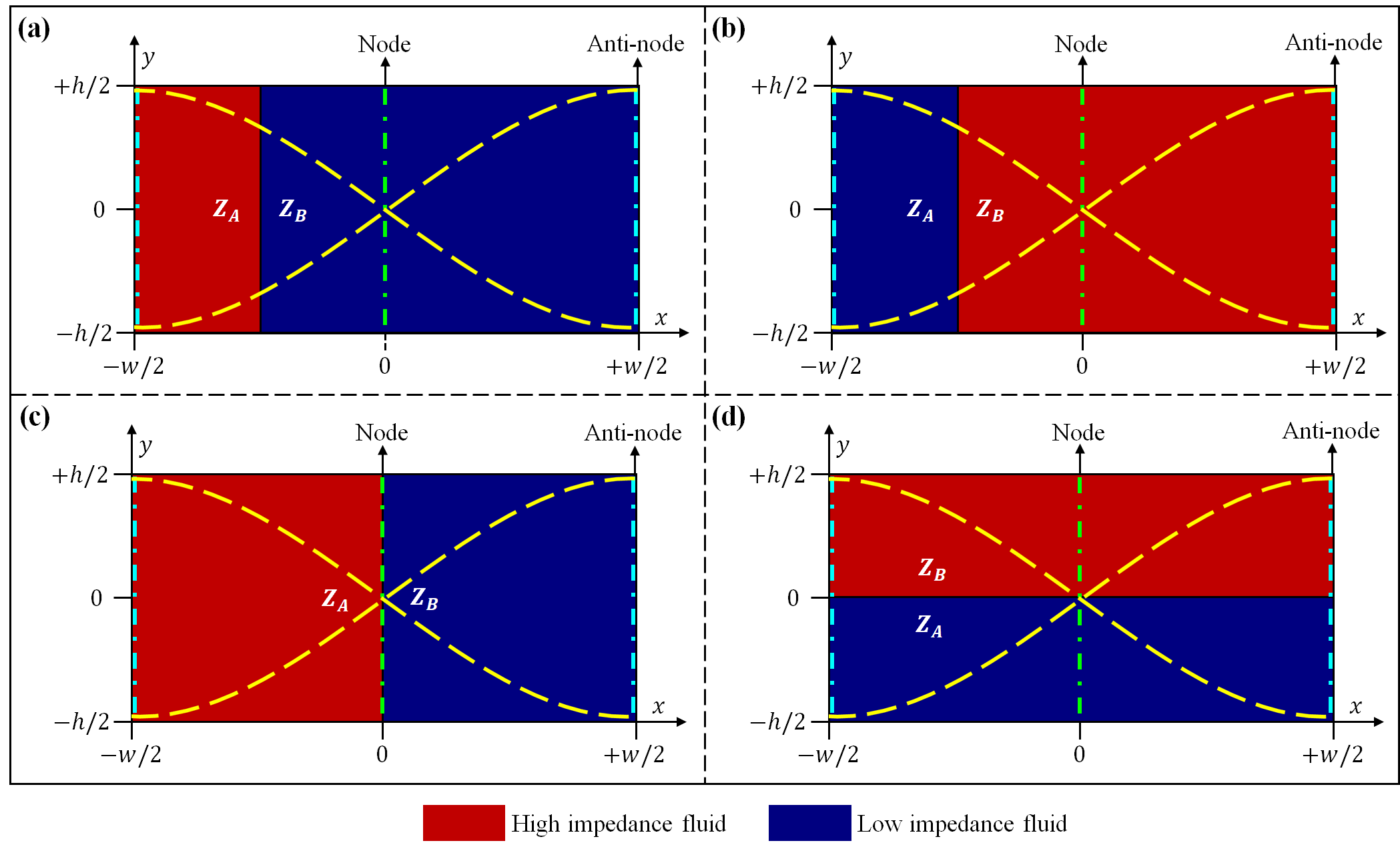}        \caption{\label{figure 1} Inhomogeneous fluids (of different impedance $Z_A$ and $Z_B$) separated by a plane interface subjected to a standing acoustic half-wave. In the absence of interfacial tension, the fluid system \textbf{(a)} is in an unstable equilibrium, \textbf{(b)} is in a stable equilibrium, \textbf{(c)} is in neutral equilibrium and \textbf{(d)} is in non-equilibrium state. In the presence of interfacial tension, the fluid system \textbf{(a)} is in a conditionally stable equilibrium, \textbf{(b)} is in stable equilibrium, \textbf{(c)} is in stable equilibrium and \textbf{(d)} is in a conditionally stable equilibrium. Note: \textbf{(d)} is not in the scope of this work.}
\end{figure} 

\subsection{Stability analysis of inhomogeneous fluids in the absence of interfacial tension\label{Sec 2.1}}
A two-dimensional fluid domain subjected to a standing acoustic half-wave in the X-direction, with two fluids separated by a sharp vertical interface as shown in figure \ref{figure 1}.a-c is considered for the stability analysis. Before beginning the analysis, it is necessary to understand the equilibrium of the system in the absence of interfacial tension. In a completely enclosed domain, a fluid initially at rest ($\textbf{\emph{V}}=0$) will remain at rest (or equilibrium) if the body force can be completely absorbed in pressure, $\textbf{\emph{f}}_{rl}=\boldsymbol{\nabla}P$ from (\ref{eq 1b}) and (\ref{eq 4}). By taking the curl of the above relation, the condition for equilibrium is given as $\boldsymbol{\nabla} \times \textbf{\emph{f}}_{rl} = 0$. Thus,
\begin{equation}
\label{eq 5}
   -\frac{E_{ac}}{z_{avg}}\left[\frac{\partial}{\partial x}\left( \cos(2k_wx)\frac{\partial z}{\partial y}\right)-\frac{\partial}{\partial y}\left(\cos(2k_wx)\frac{\partial z}{\partial x}\right)\right]  =  0.
\end{equation}  
It is clear from (\ref{eq 5}) that the given fluid configuration will be in an equilibrium state, only if the direction of the acoustic standing wave is normal to the fluid-fluid interface (the direction of the acoustic standing wave is parallel to the direction of the impedance gradient) as shown in figure \ref{figure 1}($a$-$c$). Since $\boldsymbol{\nabla} \times \textbf{\emph{f}}_{rl} \neq 0$ for the configuration shown in figure \ref{figure 1}$(d)$, it is not in equilibrium and tends to relocate to the stable configuration without any perturbations. The stability nature of the equilibrium configurations is analysed by imposing infinitesimal perturbations on the interface. 
Now we proceed to show that in the absence of interfacial tension, the configuration shown in figure \ref{figure 1}$(a)$ is in unstable equilibrium (perturbations grow), figure \ref{figure 1}$(b)$ is in stable equilibrium (perturbations decay), and figure \ref{figure 1}($c$) is in neutral equilibrium (perturbations neither grow nor decay).

The effect of viscosity is neglected in the stability analysis, as it governs only the timescale of the phenomenon and does not contribute to the stability criterion. Although the physical properties are non-uniform in an inhomogeneous system, the fluid particles considered in the flow field have constant density $\rho$, speed of sound $c$, and impedance $Z$. Thus, the material derivative of all properties is zero, which includes the incompressibility condition $(D\rho/Dt=\partial \rho/\partial t+\textbf{\emph{V}}\boldsymbol{\cdot\nabla}\rho = 0)$. By combining the incompressibility condition with (\ref{eq 1a}) and neglecting the viscosity, the governing equations (\ref{eq 1}) reduce to
\begin{subequations}
\label{eq 6}
\begin{gather}
    \label{eq 6a}
    \frac{\partial U}{\partial x}+\frac{\partial V}{\partial y}=0, \\
    \label{eq 6b}
    \rho\frac{DU}{Dt}=-\frac{\partial P}{\partial x}-\frac{E_{ac}\cos({2k_wx})}{z_{avg}}\frac{\partial Z}{\partial x}, \\
    \label{eq 6c}
    \rho\frac{DV}{Dt}=-\frac{\partial P}{\partial y}-\frac{E_{ac}\cos({2k_wx})}{z_{avg}}\frac{\partial Z}{\partial y},
\end{gather}
where $U, V$ are the X-component and Y-component of the velocity field $\textbf{\emph{V}}$. Since the body force term is a function of impedance, the below impedance relation is required for the closure.
\begin{equation}
    \label{eq 6d}
    \frac{DZ}{Dt}=\frac{\partial Z}{\partial t}+U\frac{\partial Z}{\partial x}+V\frac{\partial Z}{\partial y} = 0.
\end{equation}
\end{subequations}
Now, the flow fields are decomposed into an unperturbed zeroth-order stationary state and infinitesimal perturbations as $U=u_0+\delta u,\ V=v_0+\delta v,\ P=p_0+\delta p,\ \rho =\rho_0+\delta\rho$ and $Z=z_0+\delta z$. In this study, the variation of acoustic impedance is considered only in the X-direction (figure \ref{figure 1}($a$-$c$)), $z_0=z_0(x)$. At the stationary state $(u_0=v_0=0)$, the unperturbed zeroth-order equations become, $\frac{\partial p_0}{\partial x} = -\frac{E_{ac}\cos({2k_wx})}{z_{avg}}\frac{\partial z_0}{\partial x}$ from \ref{eq 6b}, $\frac{\partial p_0}{\partial x}=0$ from  \ref{eq 6c} and $\frac{\partial z_0}{\partial t}=0$ from \ref{eq 6d}. Using the above zeroth order relations and  neglecting the second-order terms in (\ref{eq 6}), the first-order perturbation equations governing the stability becomes
\begin{subequations}
\label{eq 7}
    \begin{gather}
    \label{eq 7a}
        \frac{\partial \delta u}{\partial x}+\frac{\partial \delta v}{\partial y}=0,\\
    \label{eq 7b}
        \rho_0\frac{\partial \delta u}{\partial t}= -\frac{\partial \delta p}{\partial x} - \frac{E_{ac}\cos({2k_wx})}{z_{avg}}\frac{\partial \delta z}{\partial x},\\
    \label{eq 7c}
        \rho_0\frac{\partial \delta v}{\partial t} = -\frac{\partial \delta p}{\partial y} - \frac{E_{ac}\cos({2k_wx})}{z_{avg}}\frac{\partial \delta z}{\partial y}, \\
    \label{eq 7d}
        \frac{\partial(\delta z)}{\partial t}=-\delta u\frac{\partial z_0}{\partial x}.
    \end{gather}
\end{subequations}
Analysing the disturbances into normal modes, the amplitude of the disturbances $\delta u,\ \delta v,\ \delta \rho,\ \delta p,$ and $\delta z$ takes the following form 
\begin{equation}
    \label{eq 8}
    A(x,y,t)=A(x)exp(ik_yy+nt),
\end{equation}
where $k_y$ is the wavenumber considered along the Y-direction. Applying the above amplitude relations in the form (\ref{eq 8}) in (\ref{eq 7}),
\begin{subequations}
\label{eq 9}
    \begin{gather}
    \label{eq 9a}
        \frac{\partial\delta u}{\partial x}+ik_y\delta v=0, \\
    \label{eq 9b}
        \rho_0 n\delta u=-\frac{\partial \delta p}{\partial x}-\frac{E_{ac}\cos{\left(2k_wx\right)}}{z_{avg}}\frac{\partial \delta z}{\partial x}, \\
    \label{eq 9c}
        \rho_0 n\delta v=-ik_y\delta p -ik_y\frac{E_{ac}\cos{\left(2k_wx\right)}}{z_{avg}}\delta z,\\
\label{eq 9d}    
    n\delta z=-\delta u\frac{\partial z_0}{\partial x}.
\end{gather}
\end{subequations}
The partial notation is dropped since the only derivatives in (\ref{eq 9}) are with respect to the $x$ coordinate. Multiplying by $ik_y$ throughout (\ref{eq 9c}) and combining with (\ref{eq 9a}) and (\ref{eq 9d}), we obtain,
\begin{equation}
\label{eq 10}
\delta p = -\rho_0 \frac{n}{{k_y}^2}\frac{d\delta u}{dx}+E_{ac}\frac{\cos(2k_wx)}{z_{avg}}\frac{\delta u}{n}\frac{dz_0}{dx}.
\end{equation}
substituting (\ref{eq 9d}) and (\ref{eq 10}) in (\ref{eq 9b}) results in,
\begin{equation}
\label{eq 11}
     \frac{d}{dx}\left(\rho_0 \frac{d\delta u}{dx} \right) - \rho_0 k_y^2 \delta u = -E_{ac}\frac{2k_w\delta u}{z_{avg}}\frac{k_y^2}{n^2}\frac{dz_0}{dx} \sin(2k_wx).
\end{equation}
Considering two uniform fluids of different impedance $Z_A$ and $Z_B$ separated by interfaces positioned at  $x_s$,
\begin{subequations}
\label{eq 12}
    \begin{equation}
    \label{eq 12a}
    z_0=z_A+(z_B-z_A)H(x-x_s),
    \end{equation}
    \begin{equation}
    \label{eq 12b}
    \frac{dz_0}{dx}=(z_B-z_A)\delta (x-x_s),
    \end{equation}
\end{subequations}
where $H(x-x_s)$ is the Heaviside step function at $x=x_s$ and  $\delta(x-x_s)$ is the Dirac’s $\delta$-function at $x=x_s$. Substituting (\ref{eq 12b}) in (\ref{eq 11}),
\begin{equation}
\label{eq 13}
    \frac{d}{dx}\left(\rho_0 \frac{d\delta u}{dx}\right) -\rho_0k_y^2\delta u= -E_{ac}\frac{2k_w\delta u}{z_{avg}}\frac{k_y^2}{n^2}\sin(2k_wx)(z_B-z_A)\delta(x-x_s).
\end{equation}
Equation (\ref{eq 13}) is the governing differential equation for the stability of inhomogeneous fluids (without interfacial tension). For a uniform region on either side of the interface(s) where there are no discontinuities in the impedance, the governing equation (\ref{eq 13}) reduces to
\begin{equation}
    \label{eq 14}
    \frac{d^2\delta u}{dx^2}-k_y^2\delta u=0.
\end{equation}
The solution of (\ref{eq 14}) is of the form $\delta u=C_1e^{k_y(x-x_s)}+C_2e^{-k_y(x-x_s)}$ where $C_1,C_2$ are constants. Since $\delta u$ must vanish at the boundaries, we can write the solution as,
\begin{subequations}
\label{eq 15}
\begin{equation}
    \label{eq 15a}
    \delta u_B = Ce^{k_y(x-x_s)}\quad\quad(x<x_s),
\end{equation}
\begin{equation}
    \label{eq 15b}
    \delta u_A = Ce^{-k_y(x-x_s)}\quad\quad(x>x_s),
\end{equation}
\end{subequations}
where the constant $C$ in (\ref{eq 15}) is chosen to ensure continuity in velocity across the interfaces. For the solution at the interface $(x=x_s)$, we integrate (\ref{eq 13}) along infinitesimal distance $(dx\approx 0)$, the second term in the left-hand side of the equation is zero and the remaining terms are, 
\begin{equation}
\label{eq 16}
\Delta\left(\rho_0 \frac{d\delta u_s}{dx}\right)= -E_{ac}\frac{2k_w\delta u_s}{z_{avg}}\frac{k_y^2}{n^2}(z_B-z_A)\int{ \left(\sin(2k_wx)\delta(x-x_s)\right)dx},
\end{equation}
where $\delta u_s$ is the value of $\delta u$ at $x=x_s$. Using,  (\ref{eq 15}) and the Dirac delta identity $\int f(x)\delta(x-a)dx = f(a)$ to solve for eigenvalue $n$ in (\ref{eq 16}). 
\begin{equation}
    \label{eq 17}
   \rho_A (- k_y \delta u_s) - \rho_B (k_y \delta u_s) = -E_{ac}\frac{2k_w\delta u_s}{z_{avg}}\frac{k_y^2}{n^2}(z_B-z_A) \sin (2 k_w x_s), 
\end{equation}
Rearranging (\ref{eq 17}), the dispersion relation $n$ for the stability problem becomes 
\begin{equation}
\label{eq 18}
    n = \sqrt{\frac{k_y}{\rho_A+\rho_B}\phi E_{ac}(z_B-z_A)\sin(2k_wx_s)}.
\end{equation}
where $\phi = 2k_w/z_{avg}$. The dispersion relation (\ref{eq 18}) establishes the acoustic stability criterion when inhomogeneous fluids (without interfacial tension) in a microchannel are subjected to a standing acoustic wave. If the eigenvalue $n$ is imaginary in (\ref{eq 18}), then the configuration is in a stable equilibrium and the configuration is in an unstable equilibrium when the eigenvalue $n$ is real. For a standing acoustic half-wave, in (\ref{eq 18}), the values of $\frac{k_{y}}{\rho_{B}+\rho_{A}}, \phi$ and $E_{a c}$ are always positive. Thus, the sign of $z_{B}-z_{A}$ (initial configuration of the fluids) and $\sin (2 k_{w} x_s)$ (relative location of the interface with respect to the standing acoustic wave) decide the nature of the eigenvalue in (\ref{eq 18}). $z_{B}-z_{A}$ is positive when high-impedance fluid is present to the right of the interface, and negative when high-impedance fluid is present to the left of the interface. $\sin (2 k_{w} x_s)$ has a negative value to the left of the node ($x_s$ is negative), a positive value to the right of the node ($x_s$ is positive) and zero when the interface coincides with the node (centre of the microchannel) or anti-node (sides of the microchannel) $(x_s = 0)$.  

\begin{figure}
    \center
    \includegraphics[width=1\linewidth]{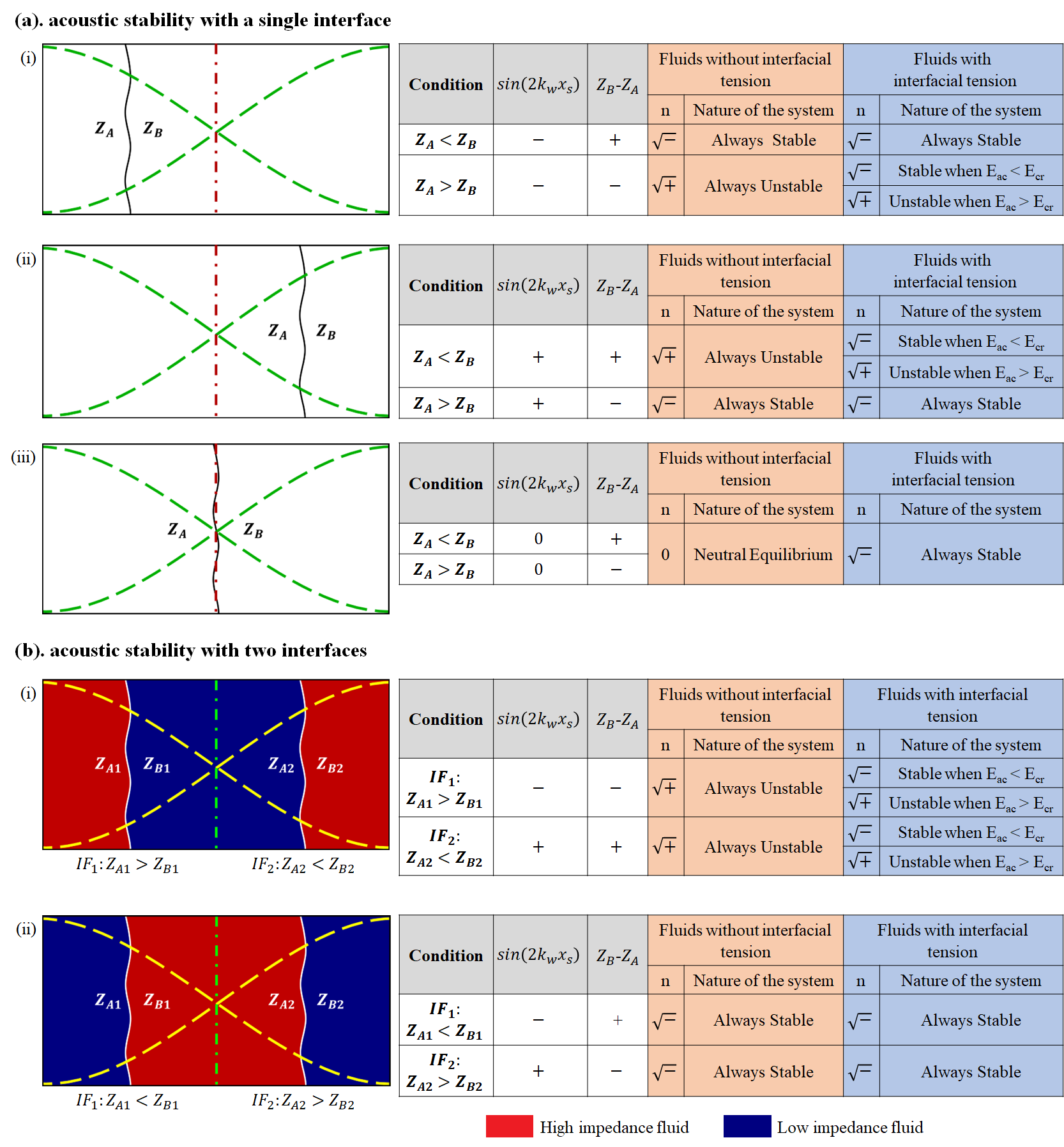}
    \caption{\label{figure 2} Different inhomogeneous fluid configurations commonly used in microfluidics and their equilibrium nature. \textbf{a)} Single interface configurations; \textbf{b)} Double interface configurations. Equations (\ref{eq 1}) and (\ref{eq 2}) are used to calculate $n$ for fluids without interfacial tension and with interfacial tension.}
\end{figure}

As per the above arguments, the inhomogeneous system in figure \ref{figure 1}($a$) is in an unstable equilibrium as eigenvalue $n$ is real, and the system in figure \ref{figure 1}($b$) is in a stable equilibrium as eigenvalue $n$ is imaginary. It can be concluded from the above discussion and figures \ref{figure 2}($a$-$i$) and \ref{figure 2}($a$-$ii$) that, a system is said to be acoustically stable (unstable) if the initial configuration of the fluids is in such a way that the low (high) impedance fluid is present at the anti-node(s) and the high (low) impedance fluid is present at the node(s). This conclusion is consistent with the demonstration of acoustic relocation of fluids within a microchannel by \citet{Deshmukh2014Jul}. For the case where the interface coincides with the node, $\sin (2 k_{w} x_s)=0$. Thus, the system is in a neutral equilibrium ($n=0$) as shown in figures \ref{figure 1}($c$) and \ref{figure 2}($a$-$iii$). The above analysis can be easily extended to an inhomogeneous system consisting of multiple interfaces. In this case, the eigenvalues evaluated at the fluid interfaces govern the nature of the system. Figure \ref{figure 2}($b$) shows the stability of two interface systems that are widely used in acoustofluidic applications. It can be seen from figure \ref{figure 2}($b$-$i$) that when high impedance fluid is at the sides (anti-nodes), the eigenvalue at both the interfaces $(IF_1$ and $IF_2)$ is real and hence the system is in unstable equilibrium. The system is in stable equilibrium in figure \ref{figure 2}($b$-$ii$), as the eigenvalue at both interfaces is imaginary.

\subsection{Stability analysis of inhomogeneous fluids in the presence of interfacial tension\label{Sec 2.2}}

Proceeding to solve for immiscible fluids, the effect of surface tension must be accounted for. The discontinuity in impedance occurring in the interfaces ($x_s$) is modelled by including the interfacial tension effects in the X momentum equation (\ref{eq 9b}) as, \citep{Chandrasekhar1961}
\begin{equation}
\label{eq 19}
    \rho_0 n\delta u = -\frac{\partial \delta p}{\partial x} - \frac{E_{ac}\cos{\left(2k_wx\right)}}{z_{avg}}\frac{\partial \delta z}{\partial x} - k_y^2\sum_{s}T\delta x_s\delta(x-x_s).
\end{equation}
where $T$ is the interfacial tension and $\delta x_s$ denotes the perturbation of the interfaces and $\frac{d}{dt}\delta x_s = \delta u_s \implies \delta x_s = \frac{\delta u_s}{n}$. The governing differential equation for stability between inhomogeneous fluids with interfacial tension is obtained similar to the case without interfacial tension, as in the previous $\S$ \ref{Sec 2.1}, 
\begin{multline}
\label{eq 20}
       \frac{d}{dx}\left(\rho_0 \frac{d\delta u}{dx} \right) - \rho_0 k_y^2 \delta u = - E_{ac}\frac{2k_w\delta u}{z_{avg}}\frac{k_y^2}{n^2}\sin(2k_wx)(z_B-z_A)\delta(x-x_s)\\+\frac{k_y^2}{n^2}\sum_{s}{k_y^2\left(T \delta u_s\right)\delta(x-x_s)},
\end{multline}
Integrating (\ref{eq 20}) across an infinitesimal distance ($dx \approx 0$) and solving for the dispersion relation $n$,
\begin{equation}
\label{eq 21}
    n = \sqrt{\frac{k_y}{\rho_1+\rho_2}\left(\phi E_{ac}(z_B-z_A)\sin(2k_wx_s)-k_y^2T\right)}.
\end{equation}
Equation (\ref{eq 21}) establishes the acoustic stability criterion when fluids with interfacial tension are subjected to a standing acoustic wave. It can be seen from (\ref{eq 21}) that the interfacial tension $(T)$ and wavenumber of the perturbation $(k_y)$ play a role in the stability of immiscible fluids. 

In the presence of interfacial tension ($T>0$), the fluid system shown in figure \ref{figure 1}($b$) is always stable, as the negative sign of $(z_{B}-z_{A}) \sin (2k_{w}x_s)$ results in an imaginary eigenvalue $n$ in (\ref{eq 21}). Whereas, for the fluid system shown in figure \ref{figure 1}($a$), the sign of  $(z_{B}-z_{A}) \sin (2k_{w}x_s)$ is positive in (\ref{eq 21}). Thus, the system is conditionally stable, and the stability is determined by the relative magnitudes of $\phi E_{a c}(z_{B}-z_{A}) \sin (2 k_{w} x_s)$ and $k_y^{2} T$. The fluid system (figure \ref{figure 1}($a$) becomes unstable ($n$ is real) if the acoustic force density $F_{rl}$ ($\phi E_{a c}(z_{B}-z_{A}) \sin (2 k_{w} x_s)$) dominates (or is greater than) the interfacial force density $F_{int}$ ($k_y^{2} T$) and becomes stable ($n$ is imaginary) if the interfacial force density dominates the acoustic force density. For the case where the interface coincides with the node ($\sin (2 k_{w} x_s)=0$ and eigenvalue $n$ is imaginary) and the system is in a stable equilibrium, as shown in figures \ref{figure 1}($c$) and \ref{figure 2}($a$-$iii$).

Now, for conditionally stable configuration, we proceed to find the minimum energy density required to relocate the fluid systems in figures \ref{figure 1}($a$) and \ref{figure 2}($b$-$i$) with interfacial tension ($T>0$). Since the interface height, $h$ is finite, this leads to the quantization of the possible modes $k_y=k_{h_n}=n\pi/h$. The minimum (critical) acoustic energy density $(E_{cr})$ required to relocate the fluid system is decided by the first conceivable mode, $k_{h_1}=k_h=\pi/h$ and the critical acoustic energy density is obtained by limiting the eigenvalue $n$ to zero in (\ref{eq 21}). Thus,
\begin{equation}
\label{eq 22}   
    E_{cr} = \frac{k_h^2Tz_{avg}}{sin(2k_wx_s)2k_w(z_B-z_A)}.
\end{equation} 
If the applied energy density $E_{ac}$ is less than the critical energy density $E_{cr}$ ($E_{a c}<E_{c r}$), the interfacial tension succeeds in stabilizing a potentially unstable configuration. The same system becomes unstable and eventually relocates to a stable configuration when $E_{a c} > E_{c r}$. The above discussions on the equilibrium nature of different inhomogeneous fluid configurations (with and without interfacial tension) are clearly summarised in figure \ref{figure 2}. 
\section{Numerical results and discussion}\label{Sec 3} 
In this section ($\S$ \ref{Sec 3}), we numerically analyze the stability of inhomogeneous fluids (with and without interfacial tension) under acoustic fields and compare them with the results obtained by the theoretical analysis in the previous section ($\S$ \ref{Sec 2}). At first, we study the stability and relocation using acoustic relocation force $\textbf{\emph{f}}_{rl}$ (\ref{eq 4}) where the acoustic energy density is assumed to be constant (as the variation of first-order pressure and velocity are not considered). We further extend the numerical analysis using the generalized acoustic body force $\textbf{\emph{f}}_{ac}$ (\ref{eq 2}) where the first-order pressure and velocity vary during relocation (thus $E_{ac}$ varies) \citep{Rajendran2022Jun}.

\begin{figure}
\center
    \includegraphics[width=1\linewidth]{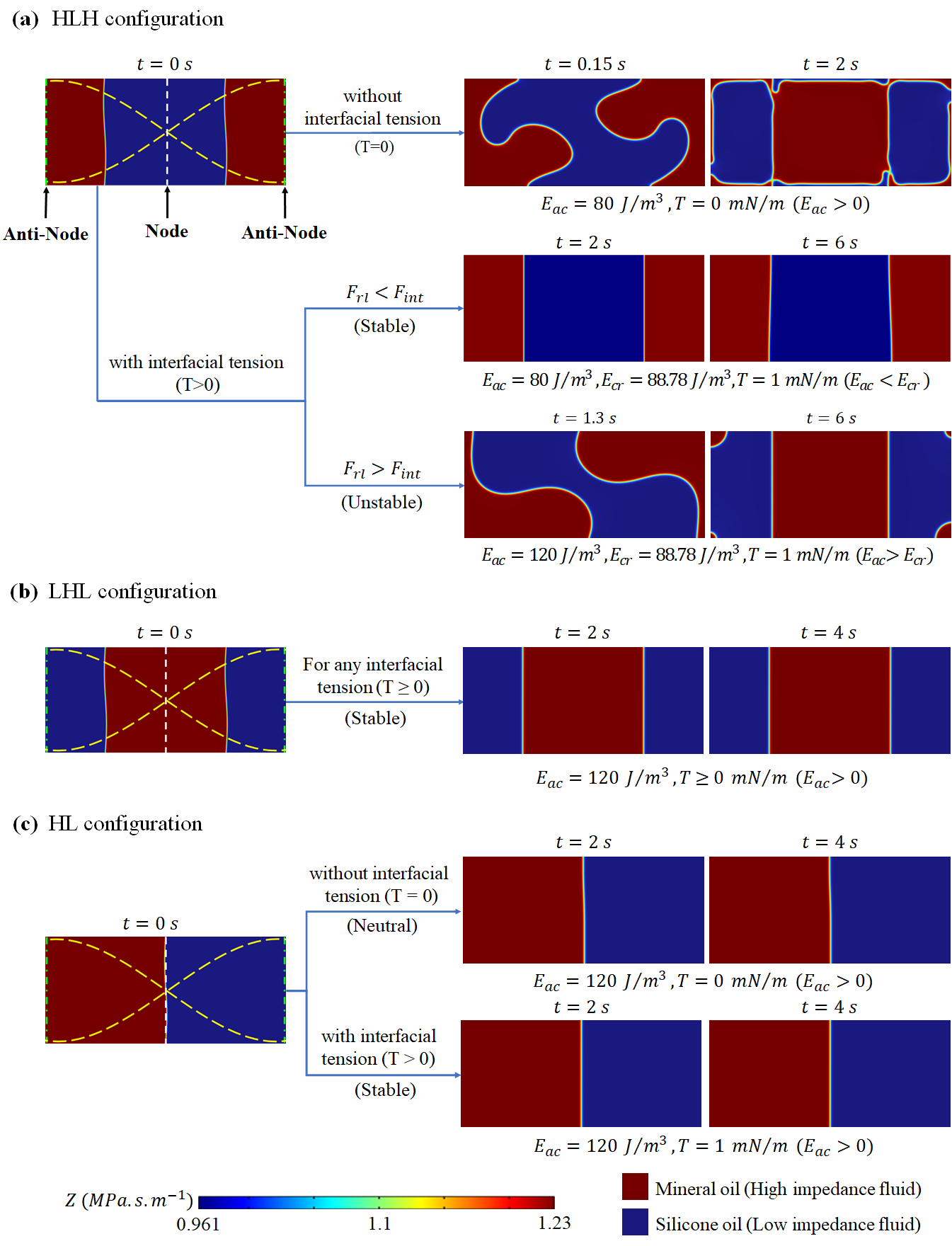}
    \caption{\label{figure 3} Stabilization and relocation of inhomogeneous fluids using simplified body force (\ref{eq 4}) with constant $E_{ac}$ - \textbf{(a)} High-Low-High (HLH) configuration, \textbf{(b)} Low-High-Low (LHL) configuration,
    \textbf{(c)} High-Low (HL) configuration}
\end{figure}

The numerical analysis is carried out on a two-dimensional fluid domain of height $h=160\ \mu m$ and width $w= 360\ \mu m$ in COMSOL Multiphysics 6.0. For this study, the fluids mineral oil $(Z=1.23\ MPa\ s/m)$ and silicone oil $(Z=0.961\ MPa\ s/m)$ are used. A mesh refinement procedure, similar to those employed by \citet{Rajendran2022Jun} is used to confirm that the numerical findings are not affected by grid size. Three different fluid configurations are considered for the study, namely, 
\begin{itemize}
    \item High-Low-High (HLH) configuration where the high impedance fluid is present at the anti-nodes (sides) and the low impedance fluid is present at the node (center) as shown in figure \ref{figure 3}($a$).
    \item Low-High-Low (LHL) configuration where the low impedance fluid is present at the anti-nodes (sides) and the high impedance fluid is present at the node (center) as shown in figure \ref{figure 3}($b$).
    \item High-Low (HL) configuration where the high impedance fluid occupies the domain to the left of the center of the microchannel and the low impedance fluid occupies the domain to the right of the center of the microchannel as shown in figure \ref{figure 3}($c$). 
\end{itemize}   

For the sake of brevity, the configurations shown in figures \ref{figure 2}($a$-$i$) (or \ref{figure 1}) and \ref{figure 2}($a$-$ii$) are not discussed explicitly as their stability and relocation are captured by HLH and LHL configurations. The Low-High (LH) configuration is also not discussed, as it would be analogous to the HL configuration. For all the analyses, the initial interface is perturbed and modelled as $x_s(y)= A_0\cos\left(\frac{2\pi}{h}y+\frac{h}{2}\right)$, where $A_0=0.01h$ is the perturbation amplitude.

\subsection{Numerical analysis of stability using constant acoustic energy density}\label{Section 3.1}

For the numerical simulations shown in \ref{figure 3}, we employ equation (\ref{eq 4}) as body force and assumed $E_{ac}$ to be constant throughout the relocation process. The boundary condition for the analysis is no slip at the walls and the pressure is constrained at a point (bottom left corner of the channel). In the absence of interfacial tension $(T=0\ mN/m)$, it is observed that for any $E_{ac}>0$, the HLH configuration undergoes relocation to a stable LHL configuration as in figure \ref{figure 3}($a$) (the simulation is shown for $E_{ac}$ $=80$ $J/m^3$). In this case, the magnitude of $E_{ac}$ only influences the timescale of the relocation process by competing with the viscosity. While, in the presence of interfacial tension, $T=1\ mN/m$, the HLH fluid configuration remained stable for all energy densities below $88\ J/m^3$, and relocation is observed for all energy densities above $89\ J/m^3$. These simulations are in close agreement with the critical acoustic energy density $E_{cr}=88.78\ J/m^3$ predicted by  (\ref{eq 22}) for mineral-silicon oil combination. Simulation results of other fluid combinations shown in figure \ref{figure 5} also agree with (\ref{eq 22}). When the applied $E_{ac}$ is just above $E_{cr}$, the fluids take a much longer time to relocate. Thus for convenience, the simulation is  shown for $E_{ac}=120\ J/m^3$ in figure \ref{figure 3}($a$).  

For LHL configuration with and without interfacial tension ($T\ge0$), for any $E_{ac}>0$, the relocation of fluid is not observed, and the system remained stable as shown in figure \ref{figure 3}($b$) (the simulation is shown for $E_{ac}$ $=120$ $J/m^3$). In the HL configuration, the node of the standing acoustic half-wave coincides with the fluid-fluid interface. Here for fluids with interfacial tension, relocation is not observed for any $E_{ac}>0$, and the fluid system remained stable (figure \ref{figure 3}($c$)). Whereas, for fluids without interfacial tension, the HL configuration is observed to be in neutral equilibrium (figure \ref{figure 3}($c$)). 
These simulation results of unstable, stable, and neutral equilibrium of inhomogeneous fluids (figure \ref{figure 3}) are in agreement with the stability criteria (from (\ref{eq 18}) and (\ref{eq 21})) that we established theoretically in $\S$ \ref{Sec 2}.

\subsection{Numerical analysis of stability using generalized body force $\textbf{f}_{ac}$\label{Section 3.2}}

\begin{figure}
    \center
    \includegraphics[width=1\linewidth]{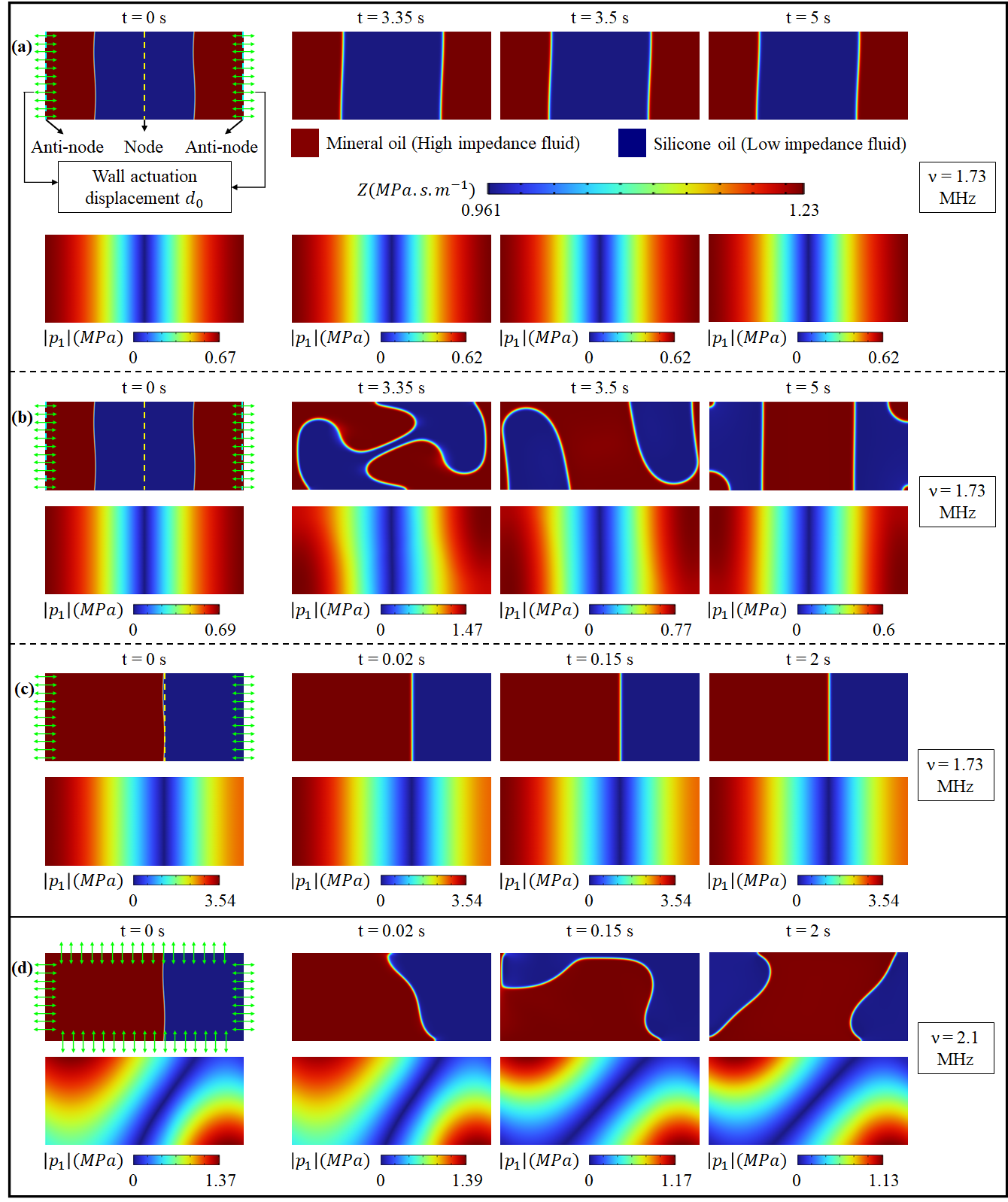}
    \caption{\label{figure 4} 
    Stabilization and relocation of inhomogeneous fluids using generalized body force $f_{ac}$ (\ref{eq 2}) along with the first-order pressure field ($|p_1|=\sqrt{Real(p_1^{\star}p_1)}$) for different fluid configurations. 1-D actuation is imposed on \textbf{(a-c)}, and 2-D actuation is imposed on \textbf{(d)}. \textbf{(a)} HLH configuration remained stable up to $E_{ac}=85.58\ J/m^3$ ($p_a = 0.67\ MPa$, $d_0$ = $0.21$ $nm$, $\nu=$  $1.73$ $M Hz$). \textbf{(b)} HLH configuration undergoes relocation above $E_{ac}=86.22$ $J/m^3$ ($p_a = 0.69\ MPa$, $d_0$ = $0.22$ $nm$, $\nu=$  $1.73$ $M Hz$). Significant  variation in $|p_1|$ during relocation is observed. \textbf{(c)} HL configuration where the fluid interface coincides with the node remained in stable equilibrium even at $E_{ac}=2334\ J/m^3$ ($p_a = 3.54\ MPa$, $d_0$ = $20$ $nm$, $\nu=$  $1.73$ $M Hz$). \textbf{(d)} Relocation of HL configuration due to 2-D wall actuation $(p_a=1.37\ MPa,\  d_0=20\ nm,\ \nu=2.1\ MHz)$
  }
\end{figure}

Thus far, in the theoretical stability analysis ($\S$ \ref{Sec 2}) as well as in the numerical simulations ($\S$ \ref{Section 3.1}), a simplified equation $\textbf{\emph{f}}_{rl}$ (\ref{eq 4}) is employed as a body force with the assumption of constant $E_{ac}$ (the amplitudes of first-order fields $p_a$ and $v_a$ do not vary during relocation). In this section, the generalized acoustic body force $\textbf{\emph{f}}_{ac}$ (\ref{eq 2}) is employed and the first-order fields required to calculate the above $\textbf{\emph{f}}_{ac}$ are obtained from the wave equations (frequency domain - see appendix \ref{Sec A1}) by actuating the channel walls at a frequency $\nu$ with a wall displacement $d_0$. There are two reasons for using generalized acoustic body force $\textbf{\emph{f}}_{ac}$: 1. To show the relocation predicted by $\textbf{\emph{f}}_{rl}$ and $\textbf{\emph{f}}_{ac}$ is approximately the same. When we use much simpler $\textbf{\emph{f}}_{rl}$ instead of the complex $\textbf{\emph{f}}_{ac}$, the first-order field equations are not required to be solved which will significantly reduce the computation time for simulation of relocation of inhomogeneous fluids. 2. To explain the previous microchannel experiments in immiscible fluid relocation \citep{Hemachandran2019Apr}. 

For one-directional (1-D) standing half-wave simulations, the sidewalls are actuated in phase at a displacement $d_0$ at a frequency $\nu$. In laminar flow equations, the boundary conditions used are no-slip at all walls, and the pressure is constrained at a point (bottom left corner of the channel). To disregard the effect of streaming, the first-order acoustic fields (see Appendix \ref{Sec A1}) are allowed to slip in the frequency domain. 

Figure \ref{figure 6}($a$), shows the HLH configuration subjected to 1-D standing half-wave by actuating sidewalls at a displacement $d_0$ of $0.21$ $nm$ and a frequency $\nu$ of $1.73\ MHz$. In this case, it is observed that the resulting pressure amplitude $P_a$ of $0.67\ MPa$ ($E_{ac}$ = $85.58\ J/m^3$), could not relocate the fluids in the HLH configuration and thus  remains stable. Whereas, when the displacement is increased to $0.22$ $nm$, the resulting pressure amplitude of $P_a=0.69\ MPa$ ($E_{ac}$ = $86.22\ J/m^3$), could relocate the HLH configuration to a stable equilibrium as shown in figure \ref{figure 6}($b$). 
From the above discussion, the critical acoustic energy density is found to be  $E_{cr}$ = $85.9\ \pm 0.32$ $J/m^3$. This value of $E_{cr}$ obtained through generalized body force $\textbf{\emph{f}}_{ac}$ is in close agreement (deviation of $3.24\%$) with the simplified relocation force $f_{rl}$ employed to derive stability criterion ($\S$ \ref{Sec 2}). 

In the case of a 1-D standing half-wave, when the interface of the fluid coincides with the pressure node ($x_s = 0$), for any $E_{ac}$, relocation is not observed using both $\textbf{\emph{f}}_{ac}$ and $\textbf{\emph{f}}_{rl}$ (figures \ref{figure 3}($c$) and \ref{figure 4}($c$) as predicted by the stability criteria (\ref{eq 21}). However, \citet{Hemachandran2019Apr} through experiments demonstrated the relocation of fluids irrespective of the location of the vertical interface $x_s$. In their experiments, the frequency employed ($2.1$ $MHz$) is far from the 1-D resonant half-wave frequency ($\nu$ = $1.6$ $MHz$ $\approx$ $c_{avg}/2w$). In our previous work \citep{Rajendran2022Jun}, we have shown that the above relocation is due to standing two-directional (2-D) acoustic wave (frequency $f$ $=$ $2.1\ MHz$ between $c_{a v g} / 2 w$ and $c_{a v g} / 2h$) as shown in figure \ref{figure 4}($d$). From figure \ref{figure 4}($d$) it is clear that the pressure node is not vertical but inclined with respect to the fluid interface owing to the 2-D actuation (all four walls are actuated at $d_0$).
The above 2-D relocation can be clearly explained by the fact that if the fluid interface and node are not perpendicular to each other, then
$\boldsymbol{\nabla} \times \textbf{\emph{f}}_{rl}$ $\neq  0$. This implies when a sufficient energy density is applied, the fluid system \ref{figure 4}($d$) will not be in equilibrium and relocation begins without imposing any perturbations unlike the other relocation discussed in this work.  

\subsection{Characterization of stable and unstable (relocation) regime\label{Section 3.3}}
\begin{figure}
    \center
    \includegraphics[width=0.7\linewidth]{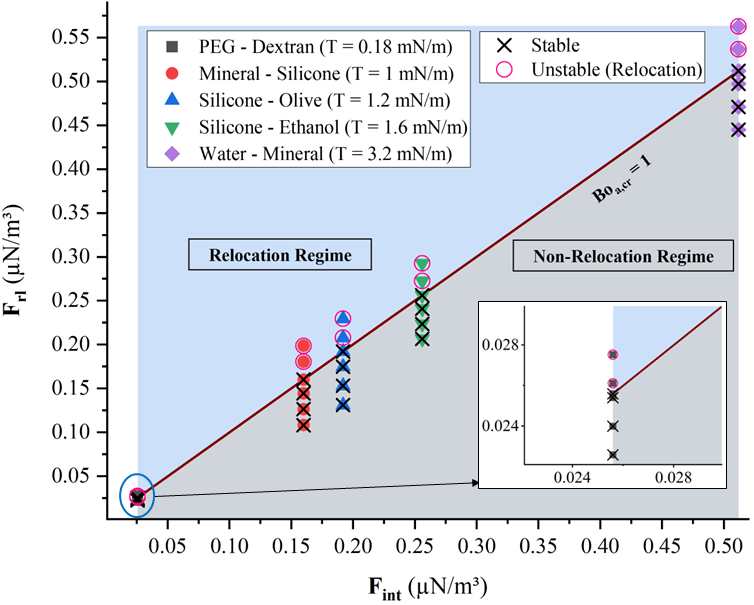}
    \caption{\label{figure 5} Characterization of relocation and non-relocation regimes of immiscible fluids using acoustic Bond number $Bo_{a}$}
\end{figure}

When the 1-D acoustic standing wave is imposed on fluids with interfacial tension, the configurations (figures \ref{figure 1}($b$), \ref{figure 2}($b$-$i$), \ref{figure 3}($a$) having high impedance fluid at the anti-node and the low impedance fluid at the node, become conditionally stable. From (\ref{eq 21}), it is evident that the stability of the above inhomogeneous fluid configurations is governed by the ratio of $F_{rl}$ and $F_{int}$, which is called as acoustic Bond number ($Bo_{a}$), given by
\begin{equation}
\label{eq 3.1}
Bo_{a}=\frac{F_{rl}}{F_{int}}=\frac{\phi E_{a c} \Delta Z \sin \left(2 k_{w} x_s\right)}{k_{h}^{2} T}   
\end{equation}
The $Bo_a$ that separates the stable and unstable region is called critical acoustic Bond number $Bo_{a,cr}$. From (\ref{eq 21})
\begin{equation}
    \label{eq 3.2}
    Bo_{a,cr}=1
\end{equation}
For $Bo_a>Bo_{a,cr}$, the above configurations become unstable (relocation occurs), and for $Bo_a<Bo_{a,cr}$ the configurations remain stable. Figure \ref{figure 5} shows the simulation results of different immiscible fluid combinations. The relocation and non-relocation regimes predicted by the simulations are in line with (\ref{eq 3.2}). It must also be noted that the fluids with higher interfacial tension require a higher acoustic force for relocation. 

\subsection{Effect of the height of the channel on stability\label{Section 3.4}}

\begin{figure}
\center
    \includegraphics[width=1\linewidth]{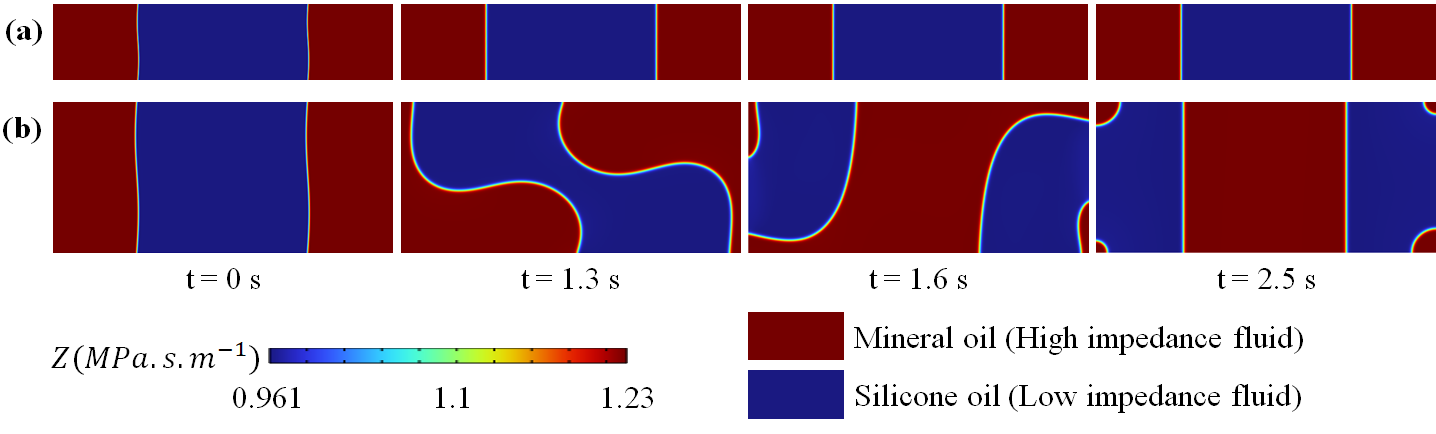}
    \caption{\label{figure 6} Effect of channel height on stability, \textbf{(a)} channel height $h= 80\ \mu m$ - no relocation is observed as applied energy density $(E_{ac} = 120\ J/m^3)$ is less than the critical energy density $(E_{cr} = 384\ J/m^3)$,  \textbf{(b)} channel height $ h = 160\ \mu m$ - relocation is observed as applied energy density $(E_{ac} = 120\ J/m^3)$ is high than the critical energy density $(E_{cr} = 88.78\ J/m^3)$. 
    This demonstrates that the interfacial tension force weakens as the height of the channel increases and thus higher the height of the channel, the lower the $E_{ac}$ required for relocation as $(E_{cr}\propto1/h^2)$.}
\end{figure}

The height of the channel $h$ plays a critical role in the stability of immiscible fluids. For a given $E_{ac}$, the increase in $h$  weakens the stabilizing effect of interfacial tension force, as analysed theoretically in $\S$ \ref{Sec 2}.  
From (\ref{eq 22}), it can be inferred that the critical acoustic energy density is inversely proportional to the square of the channel height $(E_{cr} \propto 1/h^2)$. In figure \ref{figure 6}($a$), for a microchannel of height $h=80\ \mu m$ consisting of mineral-silicone oil with interfacial tension $T=1\ mN/m$, the fluid system is stable as the applied $E_{ac}\ (120\ J/m^3$) is lower than the critical energy density ($E_{cr}=384\ J/m^3$). Whereas for $h=160\ \mu m$ and keeping the remaining parameters same, fluid relocation is observed as applied $E_{ac}\ (120\ J/m^3)$ is higher than the critical energy density $(E_{cr}=88.78\ J/m^3)$. 
 
The above discussion on the effect of channel height on acoustic relocation has high relevance in practical applications. To relocate fluids with high interfacial tension of $\mathcal{O}(10^1-10^2)$ $mN/m$, in commonly used acoustofluidic channels of height ranging from $100\ \mu m$ to $200\ \mu m$, the required $E_{ac}$ becomes $\approx \mathcal{O}(10^4)\ J/m^3$, which is very high compared to the $E_{ac}$ employed in typical acoustofluidic experiments ($\mathcal{O}(10^2-10^3)$ $J/m^3$). The equation (\ref{eq 22}) tells that the above problem can be solved by increasing the channel height as $E_{ac} \propto 1/h^2$. Hence, the depth (height) of the channel is a crucial aspect to be considered during the fabrication of an acoustofluidic microchannel for handling high interfacial tension fluids.

\section{Conclusion}
We have theoretically established the stability criteria for inhomogeneous fluids subjected to standing acoustic fields, which is consistent with the previous experimental investigations on miscible \citep{Deshmukh2014Jul}, \citep{Karlsen2016Sep}, and immiscible fluids \citep{Hemachandran2019Apr}. Numerical simulations on the same were carried out using simplified and generalized body force to understand the various parameters that contribute towards stability and relocation of fluids. However, the effect of boundary layer-driven streaming on relocation is neglected in this work, which will be addressed in an upcoming paper. The insights gained from this study can have potential applications in inhomogeneous fluid handling and particle manipulation in the field of acoustofluidics.


\appendix

\section{First-order fields in the frequency domain\label{Sec A1}}
The first-order fields (fast timescale) on the frequency domain are written as
\begin{subequations}
\label{eq A1}
\begin{equation}
\label{eq A1a}
    -i\omega\rho_1=\boldsymbol{\nabla} \cdot(\rho_0\textbf{\emph{v}}_1)
\end{equation}
\begin{equation}
\label{eq A1b}
    -i\omega\rho_0\textbf{\emph{v}}_1 = -\boldsymbol{\nabla}p_1 + \eta\nabla^2 \textbf{\emph{v}}_1 +\beta\eta\boldsymbol{\nabla}(\boldsymbol{\nabla\cdot}\textbf{\emph{v}}_1) + \textbf{\emph{f}}_{ac}
\end{equation}
\begin{equation}
\label{eq A1c}
    -i\omega\rho_0\kappa_0p_1=-i\omega\rho_1+\textbf{\emph{v}}_1\cdot\nabla\rho_0
\end{equation}
Also, combining equations (\ref{eq A1a} \& \ref{eq A1c}) we get
\begin{equation}
    -i\omega\kappa_0 p_1=-\nabla\cdot\textbf{\emph{v}}_1
    \label{eq A1d}
\end{equation}
\end{subequations}
where $p_1$ is the first-order pressure field, $\rho_1$ refers to first-order density field, $\textbf{\emph{v}}_1$ is the first-order velocity field, $\omega$ is the angular frequency, $\eta$ is the dynamic viscosity of the fluid, $\xi$ is the volume fluid viscosity, $\beta=(\xi/\eta)+(1/3)$, $f_{ac}$ is generalised body force and $Z$ is impedance. The detailed analysis of first-order and second-order fields acting on inhomogeneous fluids is given in  \citet{Rajendran2022Jun}.
\bibliographystyle{jfm}
\bibliography{jfm-instructions}

\begin{thebibliography}{25}
\expandafter\ifx\csname natexlab\endcsname\relax\def\natexlab#1{#1}\fi
\def\au#1{#1} \def\ed#1{#1} \def\yr#1{#1}\def\at#1{#1}\def\jt#1{\textit{#1}}
  \def\bt#1{#1}\def\bvol#1{\textbf{#1}} \def\vol#1{#1} \def\pg#1{#1}
  \def\publ#1{#1}\def\arxiv#1{#1}\def\org#1{#1}\def\st#1{\textit{#1}}

\bibitem[Ahmed {\em et~al.\/}(2016)Ahmed, Ozcelik, Bojanala, Nama, Upadhyay,
  Chen, Hanna-Rose \& Huang]{Ahmed2016Mar}
{\sc \au{Ahmed, Daniel}, \au{Ozcelik, Adem}, \au{Bojanala, Nagagireesh},
  \au{Nama, Nitesh}, \au{Upadhyay, Awani}, \au{Chen, Yuchao}, \au{Hanna-Rose,
  Wendy} \& \au{Huang, Tony~Jun}} \yr{2016}  \at{{Rotational manipulation of
  single cells and organisms using acoustic waves}}.  \jt{Nat. Commun.}
  \bvol{7}~(11085),  \pg{1--11}.

\bibitem[Augustsson {\em et~al.\/}(2016)Augustsson, Karlsen, Su, Bruus \&
  Voldman]{augustsson2016iso}
{\sc \au{Augustsson, Per}, \au{Karlsen, Jonas~T}, \au{Su, Hao-Wei}, \au{Bruus,
  Henrik} \& \au{Voldman, Joel}} \yr{2016}  \at{Iso-acoustic focusing of cells
  for size-insensitive acousto-mechanical phenotyping}.  \jt{Nature
  communications}  \bvol{7}~(1),  \pg{1--9}.

\bibitem[Baudoin {\em et~al.\/}(2020)Baudoin, Thomas, Sahely, Gerbedoen, Gong,
  Sivery, Matar, Smagin, Favreau \& Vlandas]{Baudoin2020Aug}
{\sc \au{Baudoin, Michael}, \au{Thomas, Jean-Louis}, \au{Sahely, Roudy~Al},
  \au{Gerbedoen, Jean-Claude}, \au{Gong, Zhixiong}, \au{Sivery, Aude},
  \au{Matar, Olivier~Bou}, \au{Smagin, Nikolay}, \au{Favreau, Peter} \&
  \au{Vlandas, Alexis}} \yr{2020}  \at{{Spatially selective manipulation of
  cells with single-beam acoustical tweezers}}.  \jt{Nat. Commun.}
  \bvol{11}~(4244),  \pg{1--10}.

\bibitem[Chandrasekhar(1961)]{Chandrasekhar1961}
{\sc \au{Chandrasekhar, Subrahmanyan}} \yr{1961} {\em {Hydrodynamic and
  hydromagnetic stability}\/}.

\bibitem[Christakou {\em et~al.\/}(2013)Christakou, Ohlin, Vanherberghen,
  Khorshidi, Kadri, Frisk, Wiklund \&
  {\ifmmode\ddot{O}\else\"{O}\fi}nfelt]{Christakou2013Apr}
{\sc \au{Christakou, Athanasia~E.}, \au{Ohlin, Mathias}, \au{Vanherberghen,
  Bruno}, \au{Khorshidi, Mohammad~Ali}, \au{Kadri, Nadir}, \au{Frisk, Thomas},
  \au{Wiklund, Martin} \& \au{{\ifmmode\ddot{O}\else\"{O}\fi}nfelt,
  Bj{\ifmmode\ddot{o}\else\"{o}\fi}rn}} \yr{2013}  \at{{Live cell imaging in a
  micro-array of acoustic traps facilitates quantification of natural killer
  cell heterogeneity}}.  \jt{Integr. Biol.}  \bvol{5}~(4),  \pg{712--719}.

\bibitem[Collins {\em et~al.\/}(2015)Collins, Morahan, Garcia-Bustos, Doerig,
  Plebanski \& Neild]{Collins2015Nov}
{\sc \au{Collins, David~J.}, \au{Morahan, Belinda}, \au{Garcia-Bustos, Jose},
  \au{Doerig, Christian}, \au{Plebanski, Magdalena} \& \au{Neild, Adrian}}
  \yr{2015}  \at{{Two-dimensional single-cell patterning with one cell per well
  driven by surface acoustic waves}}.  \jt{Nat. Commun.}  \bvol{6}~(8686),
  \pg{1--11}.

\bibitem[Deshmukh {\em et~al.\/}(2014)Deshmukh, Brzozka, Laurell \&
  Augustsson]{Deshmukh2014Jul}
{\sc \au{Deshmukh, Sameer}, \au{Brzozka, Zbigniew}, \au{Laurell, Thomas} \&
  \au{Augustsson, Per}} \yr{2014}  \at{{Acoustic radiation forces at liquid
  interfaces impact the performance of acoustophoresis}}.  \jt{Lab Chip}
  \bvol{14}~(17),  \pg{3394--3400}.

\bibitem[Friend \& Yeo(2011)]{Friend2011Jun}
{\sc \au{Friend, James} \& \au{Yeo, Leslie~Y.}} \yr{2011}  \at{{Microscale
  acoustofluidics: Microfluidics driven via acoustics and ultrasonics}}.
  \jt{Rev. Mod. Phys.}  \bvol{83}~(2),  \pg{647--704}.

\bibitem[Gautam {\em et~al.\/}(2018)Gautam, Gurung, Fencl \&
  Piyasena]{Gautam2018Oct}
{\sc \au{Gautam, Gayatri~P.}, \au{Gurung, Rubi}, \au{Fencl, Frank~A.} \&
  \au{Piyasena, Menake~E.}} \yr{2018}  \at{{Separation of sub-micron particles
  from micron particles using acoustic fluid relocation combined with
  acoustophoresis}}.  \jt{Anal. Bioanal.Chem.}  \bvol{410}~(25),
  \pg{6561--6571}.

\bibitem[Hemachandran {\em et~al.\/}(2021)Hemachandran, Hoque, Laurell \&
  Sen]{Hemachandran2021Sep}
{\sc \au{Hemachandran, E.}, \au{Hoque, S.~Z.}, \au{Laurell, T.} \& \au{Sen,
  A.~K.}} \yr{2021}  \at{{Reversible Stream Drop Transition in a Microfluidic
  Coflow System via On Demand Exposure to Acoustic Standing Waves}}.  \jt{Phys.
  Rev. Lett.}  \bvol{127}~(13),  \pg{134501}.

\bibitem[Hemachandran {\em et~al.\/}(2019)Hemachandran, Karthick, Laurell \&
  Sen]{Hemachandran2019Apr}
{\sc \au{Hemachandran, E.}, \au{Karthick, S.}, \au{Laurell, T.} \& \au{Sen,
  A.~K.}} \yr{2019}  \at{{Relocation of coflowing immiscible liquids under
  acoustic field in a microchannel}}.  \jt{Europhys. Lett.}  \bvol{125}~(5),
  \pg{54002}.

\bibitem[Iranmanesh {\em et~al.\/}(2015)Iranmanesh, Ramachandraiah, Russom \&
  Wiklund]{Iranmanesh2015Sep}
{\sc \au{Iranmanesh, Ida}, \au{Ramachandraiah, Harisha}, \au{Russom, Aman} \&
  \au{Wiklund, Martin}} \yr{2015}  \at{{On-chip ultrasonic sample preparation
  for cell based assays}}.  \jt{RSC Adv.}  \bvol{5}~(91),  \pg{74304--74311}.

\bibitem[Karlsen {\em et~al.\/}(2016)Karlsen, Augustsson \&
  Bruus]{Karlsen2016Sep}
{\sc \au{Karlsen, Jonas~T.}, \au{Augustsson, Per} \& \au{Bruus, Henrik}}
  \yr{2016}  \at{{Acoustic Force Density Acting on Inhomogeneous Fluids in
  Acoustic Fields}}.  \jt{Phys. Rev. Lett.}  \bvol{117}~(11),  \pg{114504}.

\bibitem[Karlsen \& Bruus(2017)]{Karlsen2017Mar}
{\sc \au{Karlsen, Jonas~T.} \& \au{Bruus, Henrik}} \yr{2017}  \at{{Acoustic
  Tweezing and Patterning of Concentration Fields in Microfluidics}}.
  \jt{Phys. Rev. Appl.}  \bvol{7}~(3),  \pg{034017}.

\bibitem[Karlsen {\em et~al.\/}(2018)Karlsen, Qiu, Augustsson \&
  Bruus]{Karlsen2018Jan}
{\sc \au{Karlsen, Jonas~T.}, \au{Qiu, Wei}, \au{Augustsson, Per} \& \au{Bruus,
  Henrik}} \yr{2018}  \at{{Acoustic Streaming and Its Suppression in
  Inhomogeneous Fluids}}.  \jt{Phys. Rev. Lett.}  \bvol{120}~(5),  \pg{054501}.

\bibitem[Lakshmanan {\em et~al.\/}(2020)Lakshmanan, Jin, Nety, Sawyer,
  Lee-Gosselin, Malounda, Swift, Maresca \& Shapiro]{Lakshmanan2020Sep}
{\sc \au{Lakshmanan, Anupama}, \au{Jin, Zhiyang}, \au{Nety, Suchita~P.},
  \au{Sawyer, Daniel~P.}, \au{Lee-Gosselin, Audrey}, \au{Malounda, Dina},
  \au{Swift, Mararet~B.}, \au{Maresca, David} \& \au{Shapiro, Mikhail~G.}}
  \yr{2020}  \at{{Acoustic biosensors for ultrasound imaging of enzyme
  activity}}.  \jt{Nat. Chem. Biol.}  \bvol{16},  \pg{988--996}.

\bibitem[Landau \& Lifshitz(1987)]{Landau1987Aug}
{\sc \au{Landau, L.~D.} \& \au{Lifshitz, E.~M.}} \yr{1987} {\em {Fluid
  Mechanics}\/}.  \publ{Oxford, England, UK: Pergamon}.

\bibitem[Li {\em et~al.\/}(2015)Li, Mao, Peng, Zhou, Chen, Huang, Truica,
  Drabick, El-Deiry, Dao, Suresh \& Huang]{Li2015Apr}
{\sc \au{Li, Peng}, \au{Mao, Zhangming}, \au{Peng, Zhangli}, \au{Zhou, Lanlan},
  \au{Chen, Yuchao}, \au{Huang, Po-Hsun}, \au{Truica, Cristina~I.},
  \au{Drabick, Joseph~J.}, \au{El-Deiry, Wafik~S.}, \au{Dao, Ming}, \au{Suresh,
  Subra} \& \au{Huang, Tony~Jun}} \yr{2015}  \at{{Acoustic separation of
  circulating tumor cells}}.  \jt{Proc. Natl. Acad. Sci. U.S.A.}
  \bvol{112}~(16),  \pg{4970--4975}.

\bibitem[Lu {\em et~al.\/}(2019)Lu, Martin, Soto, Angsantikul, Li, Chen, Liang,
  Hu, Zhang \& Wang]{Lu2019Feb}
{\sc \au{Lu, Xiaolong}, \au{Martin, Aida}, \au{Soto, Fernando},
  \au{Angsantikul, Pavimol}, \au{Li, Jinxing}, \au{Chen, Chuanrui}, \au{Liang,
  Yuyan}, \au{Hu, Junhui}, \au{Zhang, Liangfang} \& \au{Wang, Joseph}}
  \yr{2019}  \at{{Parallel Label-Free Isolation of Cancer Cells Using Arrays of
  Acoustic Microstreaming Traps}}.  \jt{Adv. Mater. Technol.}  \bvol{4}~(2),
  \pg{1800374}.

\bibitem[Pothuri {\em et~al.\/}(2019)Pothuri, Azharudeen \&
  Subramani]{Pothuri2019Dec}
{\sc \au{Pothuri, Charish}, \au{Azharudeen, Mohammed} \& \au{Subramani,
  Karthick}} \yr{2019}  \at{{Rapid mixing in microchannel using standing bulk
  acoustic waves}}.  \jt{Phys. Fluids}  \bvol{31}~(12),  \pg{122001}.

\bibitem[Rajendran {\em et~al.\/}(2022)Rajendran, Jayakumar, Azharudeen \&
  Subramani]{Rajendran2022Jun}
{\sc \au{Rajendran, Varun~Kumar}, \au{Jayakumar, Sujith}, \au{Azharudeen,
  Mohammed} \& \au{Subramani, Karthick}} \yr{2022}  \at{{Theory of nonlinear
  acoustic forces acting on inhomogeneous fluids}}.  \jt{J. Fluid Mech.}
  \bvol{940}.

\bibitem[Shi {\em et~al.\/}(2009)Shi, Huang, Stratton, Huang \&
  Huang]{Shi2009Dec}
{\sc \au{Shi, Jinjie}, \au{Huang, Hua}, \au{Stratton, Zak}, \au{Huang, Yiping}
  \& \au{Huang, Tony~Jun}} \yr{2009}  \at{{Continuous particle separation in a
  microfluidic channelvia standing surface acoustic waves (SSAW)}}.  \jt{Lab
  Chip}  \bvol{9}~(23),  \pg{3354--3359}.

\bibitem[Van~Assche {\em et~al.\/}(2020)Van~Assche, Reithuber, Qiu, Laurell,
  Henriques-Normark, Mellroth, Ohlsson \& Augustsson]{VanAssche2020Feb}
{\sc \au{Van~Assche, David}, \au{Reithuber, Elisabeth}, \au{Qiu, Wei},
  \au{Laurell, Thomas}, \au{Henriques-Normark, Birgitta}, \au{Mellroth, Peter},
  \au{Ohlsson, Pelle} \& \au{Augustsson, Per}} \yr{2020}  \at{{Gradient
  acoustic focusing of sub-micron particles for separation of bacteria from
  blood lysate}}.  \jt{Sci. Rep.}  \bvol{10}~(3670),  \pg{1--13}.

\bibitem[Xie {\em et~al.\/}(2020)Xie, Rufo, Zhong, Rich, Li, Leong \&
  Huang]{Xie2020Dec}
{\sc \au{Xie, Yuliang}, \au{Rufo, Joseph}, \au{Zhong, Ruoyu}, \au{Rich,
  Joseph}, \au{Li, Peng}, \au{Leong, Kam~W.} \& \au{Huang, Tony~Jun}} \yr{2020}
   \at{{Microfluidic Isolation and Enrichment of Nanoparticles}}.  \jt{ACS
  Nano}  \bvol{14}~(12),  \pg{16220--16240}.

\bibitem[Zhang {\em et~al.\/}(2020)Zhang, Tian, Bachman, Zhang \&
  Huang]{Zhang2020Mar}
{\sc \au{Zhang, Liying}, \au{Tian, Zhenhua}, \au{Bachman, Hunter}, \au{Zhang,
  Peiran} \& \au{Huang, Tony~Jun}} \yr{2020}  \at{{A Cell-Phone-Based
  Acoustofluidic Platform for Quantitative Point-of-Care Testing}}.  \jt{ACS
  Nano}  \bvol{14}~(3),  \pg{3159--3169}.

\end{thebibliography}
\end{document}